\documentclass[fleqn,12pt,twoside]{article}
\usepackage{espcrc1}

\usepackage{graphicx}
\usepackage[figuresright]{rotating}

\newcommand{\lap}{\mathrel{ \rlap{\raise.5ex\hbox{$<$}}
                    {\lower.5ex\hbox{$\sim$}}  } }

\newcommand{\fo}{$^{18}$F}
\newcommand{\fn}{$^{19}$F}
\newcommand{\nen}{$^{19}$Ne}

\newcommand{\pa}{$^{18}$F(p,$\alpha)^{15}$O}
\newcommand{\dpa}{D($^{18}$F,p$\alpha)^{15}$N}

\newcommand{\zaa}{{\it Astron. Astrophys.}}

\newcommand{\zapj}{{\it Astrophys. J.}}
\newcommand{\znp}{{\it Nucl.~Phys.}}

\newcommand{\zpr}{{\it Phys.~Rev.}}

\newcommand{\znim}{{\it Nucl.~Inst.~and~Meth.}}

\newcommand{\zmnras}{{\it MNRAS}}
\newcommand{\zsitges}{{\it Proceedings of the International Conference on 
		      Classical Nova Explosion}, Sitges, Spain, 20-24 May 2002, 
		      AIP, 2002}

\newcommand{\AmS}{{\protect\the\textfont2
  A\kern-.1667em\lower.5ex\hbox{M}\kern-.125emS}}

\title{Study of the \pa\ reaction for application to nova $\gamma$-ray emission}

\author{N.~de~S\'er\'eville\address[CSNSM]{CSNSM, CNRS/IN2P3/UPS, B\^at.~104, 91405 Orsay Campus, France},
        A.~Coc\addressmark,
        C.~Angulo\address[CRC]{CRC and FYNU, UCL, Chemin du Cyclotron 2, B-1248 Louvain-La-Neuve, Belgium},
        M.~Assun\c{c}\~ao\addressmark[CSNSM],
        D.~Beaumel\address[IPNO]{Institut de Physique Nucl\'eaire, CNRS/IN2P3/UPS, 91406 Orsay Campus, France},
        B.~Bouzid\address[USTHB]{USTHB, B.P. 32,  El-Alia, Bab Ezzouar, Algiers, Algeria},
        S.~Cherubini\addressmark[CRC],
        M.~Couder\addressmark[CRC],
        P.~Demaret\addressmark[CRC],
        F.~de~Oliveira~Santos\address{GANIL, B.P. 5027, 14021 Caen Cedex, France},
        P.~Figuera\address[LNS]{Laboratori Nazionali del Sud, INFN, Via S. Sofia, 44 - 95123 Catania, Italy},
        S.~Fortier\addressmark[IPNO],
        M.~Gaelens\addressmark[CRC],
        F.~Hammache\address{GSI mbH, Planckstr. 1, D-64291
Darmstadt,Germany}$^{,}$\addressmark[IPNO],
        J.~Kiener\addressmark[CSNSM],
        D.~Labar\address{Unit\'e de Tomographie Positron, UCL, Chemin du Cyclotron 2, B-1248 Louvain La Neuve, Belgium},
        A.~Lefebvre\addressmark[CSNSM],
        P.~Leleux\addressmark[CRC],
        M.~Loiselet\addressmark[CRC],
        A.~Ninane\addressmark[CRC],
        S.~Ouichaoui\addressmark[USTHB],
        G.~Ryckewaert\addressmark[CRC],
        N.~Smirnova\address{Instituut voor Kern en Stralingsfysika, Celestijnenlaan 200D, B-3001, Leuven, Belgium},
        V.~Tatischeff\addressmark[CSNSM],
        and
        J.-P.~Thibaud\addressmark[CSNSM]
        }

\begin{document}

\maketitle

\begin{abstract}
The \pa\ reaction is recognized as one of the most important reaction for
nova gamma--ray astronomy as it governs the early $\leq$ 511~keV
emission. However, its rate remains largely uncertain at nova
temperatures due to unknown low--energy resonance strengths. 
In order to better constrain this reaction rate, we have studied the 
one--nucleon transfer reaction, \dpa, at the CRC-RIB facility at 
Louvain~La~Neuve.
\end{abstract}

\section{INTRODUCTION}

Gamma--ray emission from classical novae is dominated, during the
first hours, by positron annihilation following the beta
decay of radioactive nuclei. The main contribution comes from the
decay of \fo\ (half--life of 110~mn) and hence is directly related
to \fo\ nucleosynthesis during the outburst\cite{Gom98,Her99,F00}. 
The \pa\ reaction is 
the main mode of \fo\ destruction and has been the object of
many recent experiments\cite{Gra00,Bar01,Bar02} but its rate
remains poorly known at nova temperatures.
The uncertainties are directly related to the unknown proton widths
of the first three \nen\ levels above proton emission threshold 
($E_x$, $J^\pi$ =
6.419~MeV, 3/2$^+$; 6.437~MeV, 1/2$^-$ and 6.449~MeV, 3/2$^+$).
The tails of the corresponding resonances (at respectively $E_R$ =
8~keV, 26~keV and 38~keV) can dominate the astrophysical factor in
the relevant energy range\cite{F00}. As a consequence of these
nuclear uncertainties, the \fo\ production in nova and the early
gamma--ray emission is uncertain by a factor of $\approx$300\cite{F00}.
This supports the need of new experimental studies to improve the
reliability of the predicted annihilation gamma--ray fluxes from
novae.

\section{EXPERIMENT}

A direct measurement of the relevant resonance strengths is
impossible due to the very low Coulomb barrier penetrability.
Hence, we used an indirect method aiming at determining
the one--nucleon spectroscopic factors in the analog levels
of the mirror nucleus (\fn) by the neutron transfer reaction
D($^{18}$F,p)$^{19}$F. Assuming the equality of spectroscopic 
factors in analog levels, the proton width can be deduced from 
the calculated single particle widths\cite{Ser02a}.
The experiment has been carried out at the {\it Centre de Recherche du
Cyclotron} of Louvain-La-Neuve (Belgium) by bombarding deuteriated
polypropylene (CD$_2$) targets ($\approx$ 100 $\mu$g/cm$^2$) with a 
14~MeV \fo\ radioactive beam (average intensity of $2.2 \times 10^6$
\fo\ s$^{-1}$ on target and  $^{18}$O / \fo\ $\le$ 10$^{-3}$ purity). 
The experimental setup has been described
elsewhere\cite{Ser02a} and consists of two silicon multistrip 
detectors LAMP and LEDA\cite{Dav00}. LAMP covers backwards 
laboratory angles and was used to measure the angular distribution 
of the protons, whereas LEDA covers forward laboratory angles. The 
levels of astrophysical interest are situated high above the alpha 
emission threshold (at 4.013~MeV) and mainly decay through 
$^{19}$F$^*\to^{15}$N+$\alpha$. Hence, to reduce background, we 
required coincidences between a proton in LAMP and a $^{15}$N in LEDA.

\section{DATA ANALYSIS AND RESULTS}

\begin{figure}[ht]
  \centering
  \includegraphics[width=14cm]{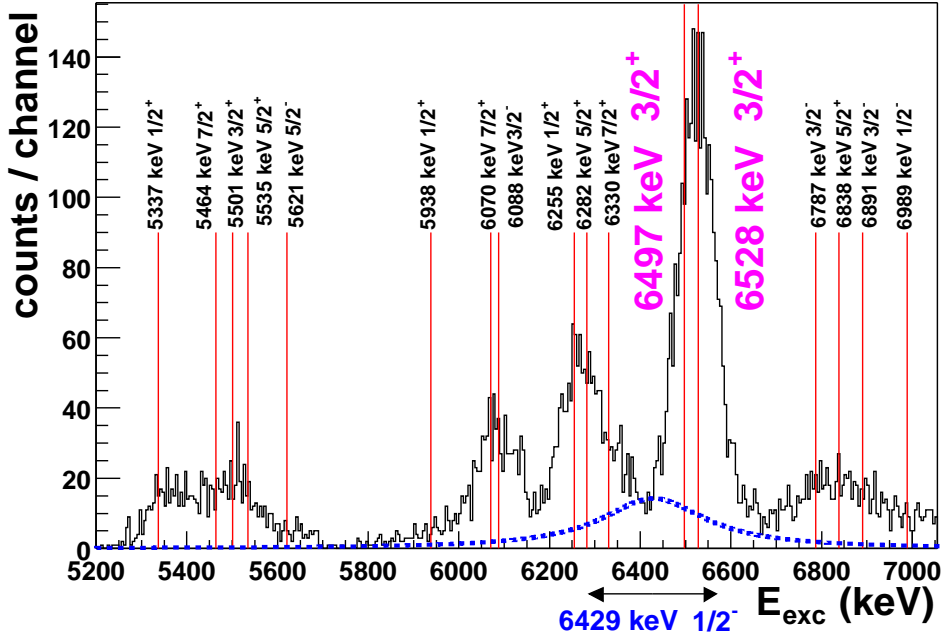}
  \caption{Reconstructed \fn\ excitation energy spectrum for coincidence 
  events (65\% of the total statistics), showing the two 3/2$^+$ levels of 
  astrophysical interest around 6.5~MeV. Vertical lines show the known
  position of the \fn\ levels populated with low transferred angular
  momentum ($l \leq 2$). The dashed line shows the broad 1/2$^-$ level for 
  $S = 0.15$ (see text).}
  \label{f:spect}
\end{figure}

Due to the kinematics only p and $\alpha$ from the
D($^{18}$F,p)$^{19}$F and D($^{18}$F,$\alpha)^{16}$O reactions can reach
LAMP. 
When considering the coincidence events each reaction
corresponds to a different zone in the ($E_{LAMP} \times
E_{LEDA}$) spectrum and hence can be identified. A further TOF
selection in LEDA is done to discriminate the p--$^{15}$N from the
p--$\alpha$ coincidences. The excitation energy of the decaying
\fn\ levels can be kinematically reconstructed from the energies
and angles of the detected protons. This
has been done by taking into account the energy loss of the \fo\ and
the protons in the CD$_2$ target as well as the energy loss of the
protons in the LAMP dead layer. The corresponding spectrum is shown
in Figure~\ref{f:spect}. The resolution is not sufficient to separate
the various levels but the two 3/2$^+$ levels of interest at 6.497
and 6.528~MeV (the analogs of the 3/2$^+$ levels in \nen) are well
separated from the other groups of levels. The spectrum is limited at
low excitation energy because of the p--$^{15}$N coincidence condition
and at high excitation energy by the electronic threshold. However, if
the coincidence condition is removed, the \fn\ spectrum extends to the
ground state and can be used for the energy calibration using
isolated peaks. However the uncertainty obtained on the excitation energy 
around the 6.5~MeV peak (two 3/2$^+$ levels) prevents a reliable
extraction of the individual contribution of these two levels.

\begin{figure}[ht]
  \centering
  \includegraphics[width=13cm]{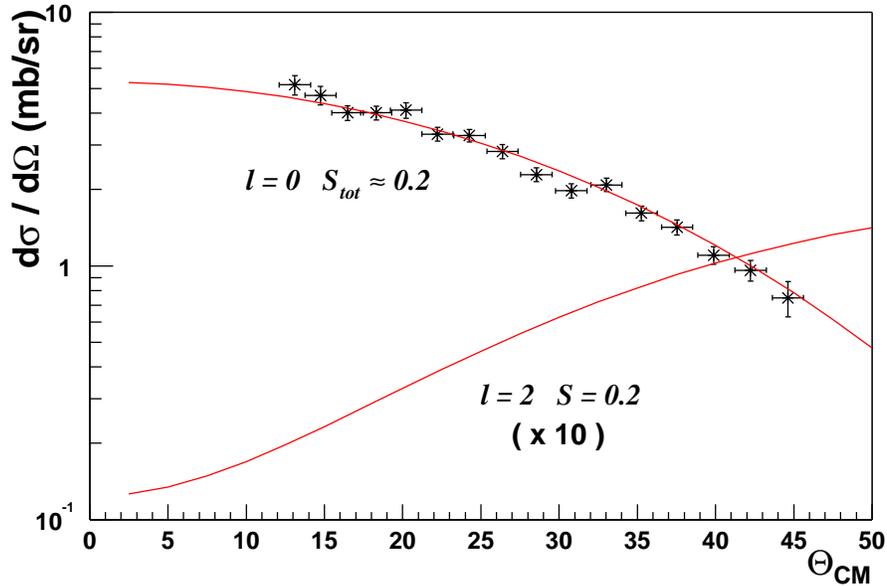}
  \caption{Comparison between the experimental angular distribution
  of the 6.5~MeV peak
  and DWBA calculations for different transferred angular momentum. The
  vertical error bars are only statistical whereas the horizontal ones
  are the angular width of each strip as seen from the target.}
  \label{f:dsig}
\end{figure}

Making a selection on the 6.5~MeV peak of the coincidence spectrum  
(Figure~\ref{f:spect}), we obtain the preliminary angular distribution shown 
in Figure~\ref{f:dsig}.
The coincidence efficiency for each strip (angle) is determined from a 
Monte-Carlo simulation taking an isotropic angular distribution for the 
$\alpha$--decay of \fn\ as a first approximation. We used the elastic 
scattering of the \fo\ beam (detected in LEDA) on the $^{12}$C of the 
target for normalization. 

The solid lines in Figure~\ref{f:dsig} correspond to theoretical
DWBA calculations with nuclear potential from Ref.~\cite{Lop64} for
different transferred angular momentum ($l = 0, 2$). This nuclear potential 
has been determined by studying a similar neutron transfer reaction 
$^{19}$F(d,p)$^{20}$F at the same center-of-mass energy (subcoulomb 
transfer) where no compound nucleus component was seen\cite{Lop64}. The 
comparison between the shapes of the theoretical and experimental
angular distributions indicates a predominant $l = 0$ 
transfer for the sum of the contributions of the two 3/2$^+$ levels. The 
value obtained for the total spectroscopic factor is $S_{tot} \approx 0.2$. 
Although no peak is seen in Figure~\ref{f:spect} corresponding to the 
1/2$^-$ level  ($E_x$ = 6.429 MeV, $\Gamma$ = 280 keV) due to its large 
total width, it is possible to derive an upper limit for the spectroscopic 
factor of $S \lap 0.15$ assuming an $l = 1$ transfer.

The important consequence of these preliminary values is that the
contribution of these resonances to the destruction rate of \fo\
{\em cannot} be neglected but remains compatible with the nominal
rate\cite{F00}. The
rate uncertainty is reduced by a factor of $\approx 5$ in the temperature
range of novae, mainly due to the reduced contribution of the 1/2$^-$
level. The impact on the rate for the two 3/2$^+$ ($S_{tot} \approx 0.2$) is
more important at low temperature\cite{Ser02a}.

\section{ACKNOWLEDGMENTS}
  This work has been supported by the European Community-Access to Research
  Infrastructure action of the Improving Human Potential Programme, contract
  N$^{o}$ HPRI-CT-1999-00110.

\end{document}